\documentstyle[12pt, epsf]{article}
\input psfig.tex
\def\lsim{\lower.5ex\hbox{$\; \buildrel < \over \sim \;$}}
\def\gsim{\lower.5ex\hbox{$\; \buildrel > \over \sim \;$}}
\begin{document}
\pagenumbering{arabic}
\title{Propagating Oscillatory Shock Model  for QPOS in GRO J1655-40 During the March 2005 Outburst\\
{\large Sandip K. Chakrabarti$^{1,2}$, A. Nandi$^2$, D. Debnath$^2$, R. Sarkar$^2$ and B.G. Datta$^2$}\\
\bigskip
{\small 1. S.N. Bose National Centre for Basic Sciences, JD-Block, Sector-III, Salt Lake, Kolkata 700098 \\
\smallskip
2. Centre for Space Physics, 43 Chalantika, Garia Station Road, Kolkata 700084}}

\maketitle

\begin{abstract}

GRO 1655-40, a well known black hole candidate, showed renewed X-ray activity in March 2005 after being dormant
for almost eight years. It showed very prominent quasi-periodic oscillations. We analysed the data of 
two observations in this {\it Rapid Communication}, one taken on  March 2nd, 2005 and  
the other taken on the March 11th, 2005.  On March 2nd, 2005 the shock was weak
and the QPO was seen in roughly all energies. On March 11th, 2005 the power density spectra showed that 
quasi-periodic oscillations  (QPOs) were exhibited in harder X-rays. On the first day, the QPO was seen at $0.13$Hz
and on the second day, the QPO was seen at $\sim 6.5$Hz with a spectral break at $\sim 0.1$Hz. 
We analysed the QPOs for the period 25th Feb. 2005 to 12th of March, 2005 and showed that 
the frequency of QPO increased monotonically from $0.088$Hz to $15.01$Hz. This agrees well if the 
oscillating shock is assumed to propagate with a constant velocity. On several days we
also noticed the presence of very high frequency QPOs and for the first time we detected 
QPOs in the $600-700$Hz range, the highest frequency range so far reported for any black hole candidate.

\end{abstract}



\noindent {Received 11th June, 2005; Accepted 4th of July, 2005; Published 1st August, 2005}

\noindent Published in Indian J. Phys. v. 79(8), p. 1-5, 2005\\

\medskip

\section{Introduction}

The well known  galactic microquasar  GRO J1655-40, which is about $3$kpc away from us [1],
and had been dormant since 1997 has suddenly come to life in March, 2005 [2]. The  RXTE X-ray count
has started rising on a daily basis. In this {\it Rapid Communication}, 
we shall examine its behaviour in two of the observations, one taken at the early stage of 
the outburst on the March 2nd, 2005 (Obs. ID 90428-0101-09) and the other taken when the outburst is 
well under way on the 11th of March, 2005 (Obs. ID: 91702-01-02-00G). The shocks may be generated 
due to the appearance of a new wave of matter and they steepen as the shock propagates
closer to the black hole. The oscillation of the shock model explains all the aspects of 
the quasi-periodic oscillations [3-6] and the QPO frequency is found to be generally related to 
the infall time $t_{if}=\frac{1}{R}\frac{1}{x_s(x_s-1)}$, where, $R$ is the shock compression
ratio $\sim 4$ and $x_s$ is the instantaneous location of the shock.
In case the shock propagates towards the black hole, the frequency is expected to 
rise monotonically. This is would be of interest if we can show (a) that the QPOs are 
indeed from shock oscillations and (b) shock frequency increases monotonically according
to the shock propagation model. In Chakrabarti \& Manickam [5] and Rao et al. [6], it was shown 
that another black hole candidate GRS 1915+105 had exhibited  QPOs  where 
shock oscillation model was  found to be valid. During the 1996/1997 outburst of 
GRO J1655-40, Remillard et al. [7] also reported energy dependence of high frequency QPOs
although no satisfactory model was discussed which could explain these behaviours. In this paper,
we also report the possible presence of a very high frequency QPO at $\sim 600$Hz, the highest 
known QPO for any black hole candidate to date.

In the next Section, we present a brief introduction of the accretion processes around black holes.
In \S 3, we present the results of the observations. We show that the nature of the hardness
ratios have changed qualitatively and quantitatively in between these two days. We also plot the
power density spectra of the photons at different energy. In \S 4 we make concluding remarks.

\section{Brief description of the black hole accretion}

Detailed descriptions of black hole accretion are given in several review articles 
and in several recent articles [8-10] and
we do not repeat them here. Briefly speaking, a black hole accretion is necessarily sub-Keplerian, i.e.,
the specific angular momentum of matter is less than the Keplerian value because of the fact that the
flow must satisfy the inner boundary condition on the horizon where matter must be supersonic
and its velocity must be equal to the velocity of light. Given that matter moves very rapidly,
the infall time becomes very low compared to any other time scale, specifically, the viscous time
scale in which angular momentum of matter is transported outward. Because angular momentum remains
virtually constant, the centrifugal barrier becomes so strong that matter almost halts at the barrier.
This is known as the shock wave at which matter undergoes a super-sonic to sub-sonic transition. The location
of the shock may form anywhere between $\sim 10$ to $\sim 1000 r_g$ (Henceforth, our unit of distance will
be the Schwarzschild radius of the central black hole: $r_g=2GM/c^2$, where
$G$ and $c$ are the  gravitational constant and the velocity of light respectively and $M$ is the mass of the 
black hole) depending on the specific angular momentum of the flow, with a typical value of around $10-15$. 
A standing shock forms where the Rankine-Hugoniot conditions are satisfied. When the conditions are 
not satisfied the shock starts oscillating depending on the exact parameters of the flow. 

It has been argued in Molteni, Sponholz and Chakrabarti [3] that the oscillating shocks will cause 
QPOs in X-rays. In Chakrabarti \& Manickam [5] and Rao et al. [6] it was shown that the
microquasar GRS 1915+105 indeed exhibited QPOs which are correctly modeled by shock-oscillations. They also
showed that while the soft X-rays do not show much oscillations, hard X-rays do. This is because the
post-shock region (otherwise known as the CENBOL -- CENtrifugal pressure dominated BOundary Layer) produces
hard X-rays by inverse-Comptonizing the intercepted soft X-rays emitted from the pre-shock flow
and also participates in oscillations more strongly than the pre-shock region [11]. 
 
\section {Results and discussions}

We analysed the data of the recent outburst using the PUBLIC data of RXTE satellite. We use the 
observation IDs 90428-01-01-09 (March 2nd, 2005) and 91702-01-02-00G (March 11th, 2005). We used only PCU2
as other PCUs are not very stable. The data analysis techniques have been reported elsewhere [5-6, 9-10]. In 
order to quantify the nature of the spectra, it is customary to plot the hardness ratios. We extracted the 
photon counts in three bins with Channel numbers 0-6 (0-2.87keV), 7-23 (2.87-9.81keV) and 24-249 (9.81-117keV)
respectively and saved then in $1$s time bins.  The ratios $R_1= Channel(7-23)/ Channel(0-6)$ 
and $R_2= Channel(7-23)/Channel(24-249)$ will then give an idea about how the 
spectrum is changing with time. In Fig. 1 we show the variation of $R_2$ as a function of $R_1$
for both the observations (marked). We note that though the observations were only 10 days apart, the 
hardness ratios behave completely differently. On March 2nd, the ($R_1$, $R_2$) is centered around
$\sim (10,1)$ indicating that the spectra was dominated by both the $ \sim 3-10$keV photons
and by $>10$keV photons. Low energy ($<3$keV) photons were not significantly present. 
On the other hand, on March 11th, 2005, we see that the ($R_1$, $R_2$) is centered around $\sim (9, 5.5)$ 
indicating that the spectra is dominated {\it only} by $3-10$keV photons and neither
the lowest, and nor the highest energy photons were significantly present. 

We now show that not only the hardness ratios, but the timing characteristics of the spectra
are very much different.  In Fig. 2, we show the power density spectra (PDS) in these
three energy bins. In the uppermost panel, which is drawn for $0-2.87$keV photons, there is a very weak and noisy 
QPO feature at $ \sim 0.13$Hz. In the middle panel, this feature is more prominent but the power is lower. 
The QPO frequency is at $0.13$Hz with an width $0.03$Hz. In the lower panel, the 
QPO feature is still present and possibly sharper, but the data is noisy and the power is definitely low.
This is to be contrasted with the energy dependence of the high frequency QPO seen on the 
March 11th, in which the QPO frequency is higher ($\sim 6.5$Hz) and the power
increased progressively at higher energy channels, thus providing a
convincing proof that the harder photons  emitted from the post-shock region 
are participating in the QPO process. Comparing the behaviour in Fig. 2 and Fig. 3
we believe that the shock was weaker originally and it steepened as
the frequency rises.

In order to show the propagation of the shock, in Fig. 4 we plot the daily variation of the QPO frequency 
between 25th of February, 2005 when the observation of significant 
photon count started till 12th of March, 2005 beyond which 
the QPO disappeared in this episode.  Superposed on the plot, we presented a theoretical plot which assumed that the 
shock is propagating with a constant velocity $v_0$. Hence the time measured from observation of 25/2/05 when the
shock was at $x_s=x_{s0}$ is given by $t=(x_{s0}-x_s)/v_0$. The frequency of QPO is already mentioned to 
be $\nu_s = \frac{\beta}{x_s (x_s-1)^{1/2}}$, where, $\beta \sim 1/R\sim 0.25$ is the inverse of the shock
strength $R$. The observed frequency $\nu_{s0}=0.088$ at $t=0$
directly gives $x_{s0}=1200$ as the launching radius.
The constraint on the disappearance of the QPO on the latter half of the 12th of March, 2005,
($\sim 15.6$d from the beginning of launching), gives  a very 
slow rate of shock propagation and the velocity is $v_0 \sim 1870$cm s$^{-1}$.

In the earlier outburst, Remillard et al. [7] also reported high frequency QPOs (at $300$Hz
and $450$Hz) observed in this object. Since as we understand it, the QPO frequency 
rises as the flow comes closer to the black hole, we expect to find QPOs at very high frequencies as
well. In Fig. 5. we show the plot of the PDS for the  Obs-Id: 90428-01-01-00 (25/2/05). 
We concentrated only at high ($>100$Hz) frequencies.
We find prominent power at $\sim 689$Hz. This is totally absent in the low energy. 
To our knowledge, signatures of such high frequency QPOs have never been reported for any black hole candidates. 

\section{Concluding Remarks}

In this Communication, we analyzed  in detail two observations of RXTE data obtained during the 
most recent outburst of March, 2005. 
Our conclusion is that at the initial stage of the outburst, the  shock was perhaps weaker
and the energy dependence of the power density was not very strong. However
as the shock propagates, it also steepens and the energy dependence becomes
stronger -- harder energy photons show stronger and narrow band QPOs. 
We also computed the  daily variation of the QPO frequency and fitted the daily variation of the 
observed frequency and found that the result agrees very well with the propagating shock 
wave model which is slowly moving towards the black hole.
We find that on the last day on which the QPO was seen at $\nu_s\sim 15.01$Hz, the 
shock was merely 38.7 Schwarzschild radius away from the black hole, though it
started from a distance of $1200$ on the 25th of February, 2005. By $15.6$ days, the
QPO and the shock disappeared completely which we interpret to be the disappearance of the
shock behind the event horizon. However, weak shocks and perturbations
continue to be advected. We caught a few such passages of shocks
close to the black hole. We find indications that there may be 
QPOs in hard X-rays in $600-700$Hz range. For instance, we find $\sim 689$Hz on 25/2/05.
If we assume that our propagating shock model is 
correct, then, $\nu_{QPO} \sim 689$Hz corresponds to an emission from $r=3.36$. 
For a Schwarzschild black hole, since shocks are expected to form farther than the 
'O' type sonic point and thus farther from $x=6$ [8], we believe that the black hole
could be a Kerr black hole. On the other hand we are dealing with a `propagating shock',
hence some of the conclusions regarding a `steady' shock need not be valid. In any case,
for a non-rotating black hole, the highest QPO frequency we can have is of $\nu_{max}=835$Hz
(at $x=3$). Unless we observe $\nu >\nu_{max}$ the case for a Kerr black hole need not be compelling.

We thank NASA/GSFC for putting the data in usable form in their website legacy.gsfc.nasa.gov.
This work is supported in part by ASTROSAT grants from TIFR (AN and RS) and CSIR/JRF
grant (DD).

\newpage

\bibliographystyle{plain}
{}

\newpage 

\begin{figure}
\vbox{
\vskip 0.0cm
\centerline{
\psfig{figure=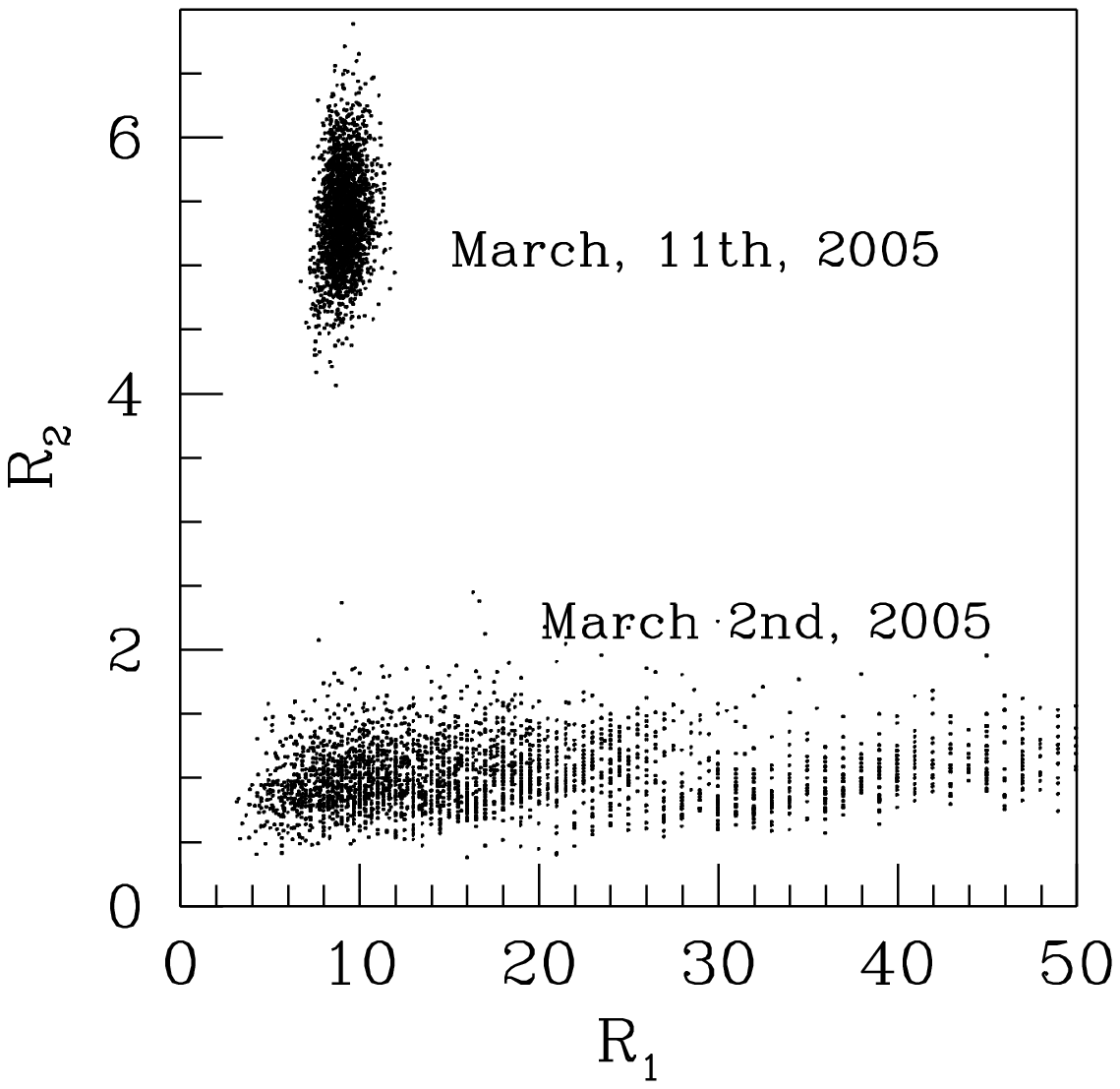,height=14truecm,width=14truecm}}}
\noindent {\small {\bf Fig. 1:}
Plots of hardness ratios $R_1 = Channel(7-23)/Channel(0-6)$ vs. $R_2=Channel(7-23)/Channel(24-249)$ 
of $1$s bin lightcurves on the (a) March 2nd, 2005 and on (b), March 11th, 2005 respectively. 
In (a), the intermediate ($2.87-9.81$) and high ($>9.81$) energy photons are dominant while in (b),
only  intermediate ($2.87-9.81$) energy photons are dominant. }
\end{figure}

\newpage

\begin{figure}
\vbox{
\vskip 0.0cm
\centerline{
\psfig{figure=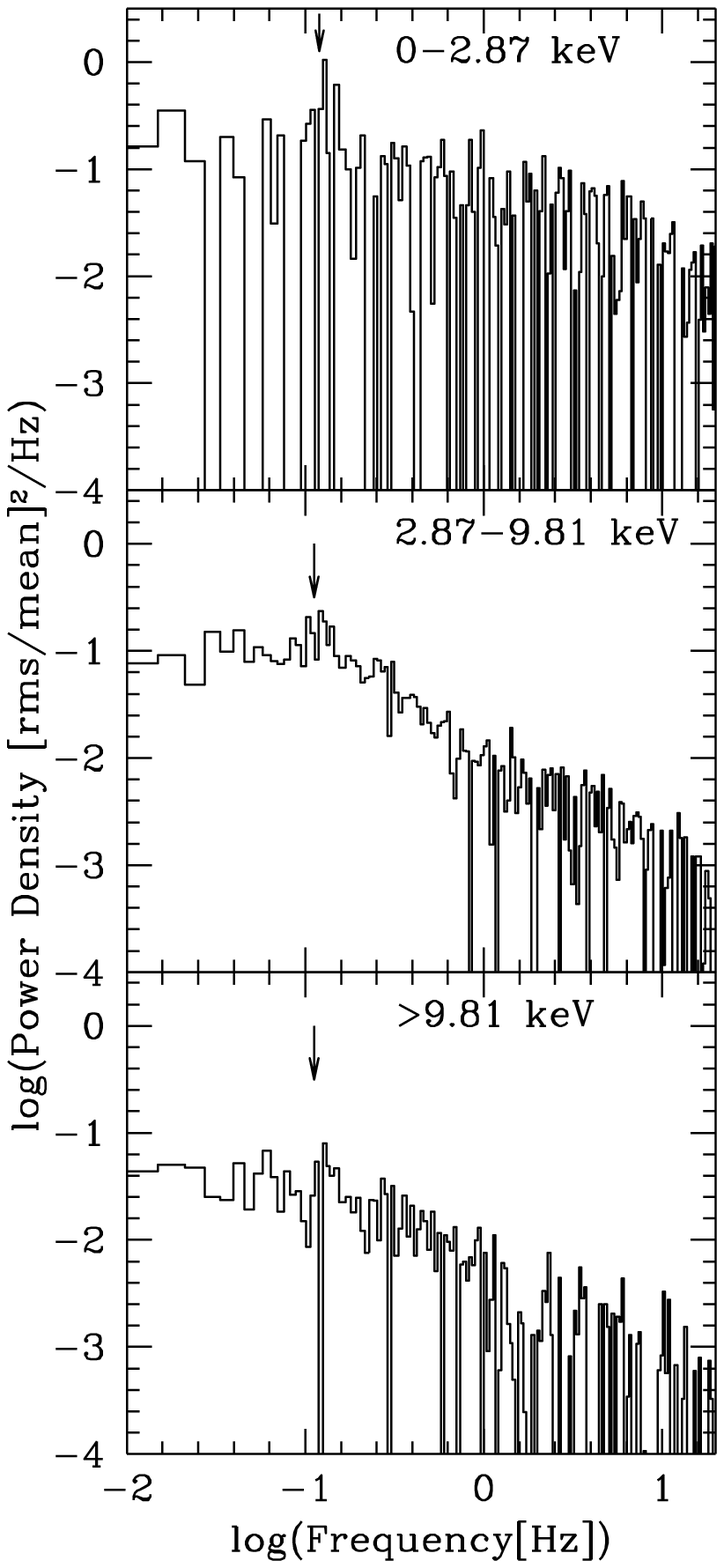,height=14truecm,width=14truecm}}}
\noindent {\small {\bf Fig. 2:}
Fig. 2: Power density spectra (PDS) of the (a) low ($0-2.87$keV), (b) intermediate ($2.87-9.81$keV)
and (c) high ($>9.81$keV) photons of March 2nd, 2005 (Obs ID:90428-0101-09)
showing that the QPOs are exhibited by higher energy photons coming from the
post-shock region. In (c) the photon counts are very low and the QPO is very noisy, though sharper.}

\end{figure}

\newpage

\begin{figure}
\vbox{
\vskip 0.0cm 
\centerline{ 
\psfig{figure=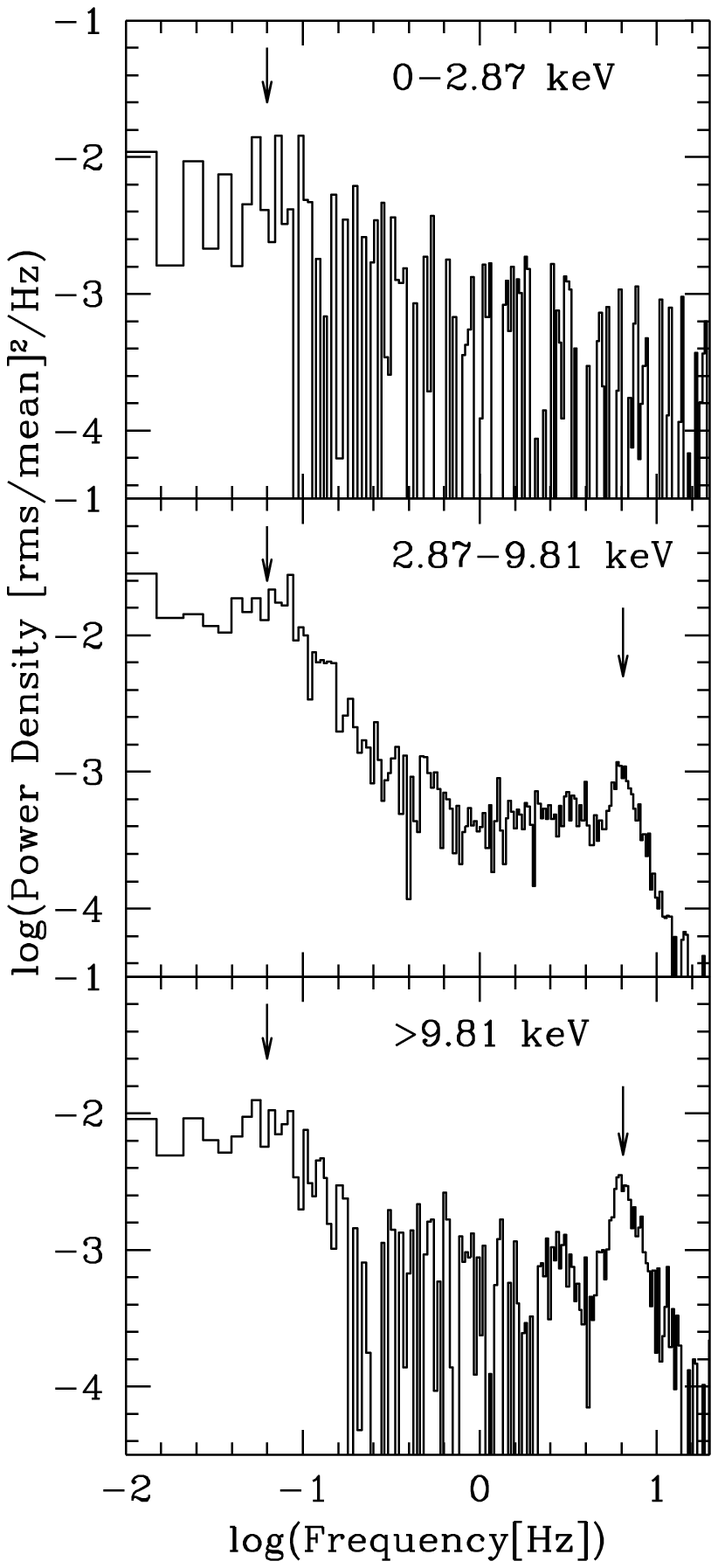,height=14truecm,width=14truecm}}}
\noindent {\small {\bf Fig. 3:}
Fig. 3: Power density spectra (PDS) of the (a) low ($0-2.87$keV), (b) intermediate ($2.87-9.81$keV)
and (c) high ($>9.81$keV) photons of March 11th, 2005 (Obs ID:91702-01-02-00G)
showing that the QPOs are exhibited by higher energy photons coming from the 
post-shock region. The power at $\nu_{QPO} \sim 6.5$Hz in (c) is three times more compared to that in (b).}

\end{figure} 

\newpage

\begin{figure}
\vbox{
\vskip 0.0cm 
\centerline{ 
\psfig{figure=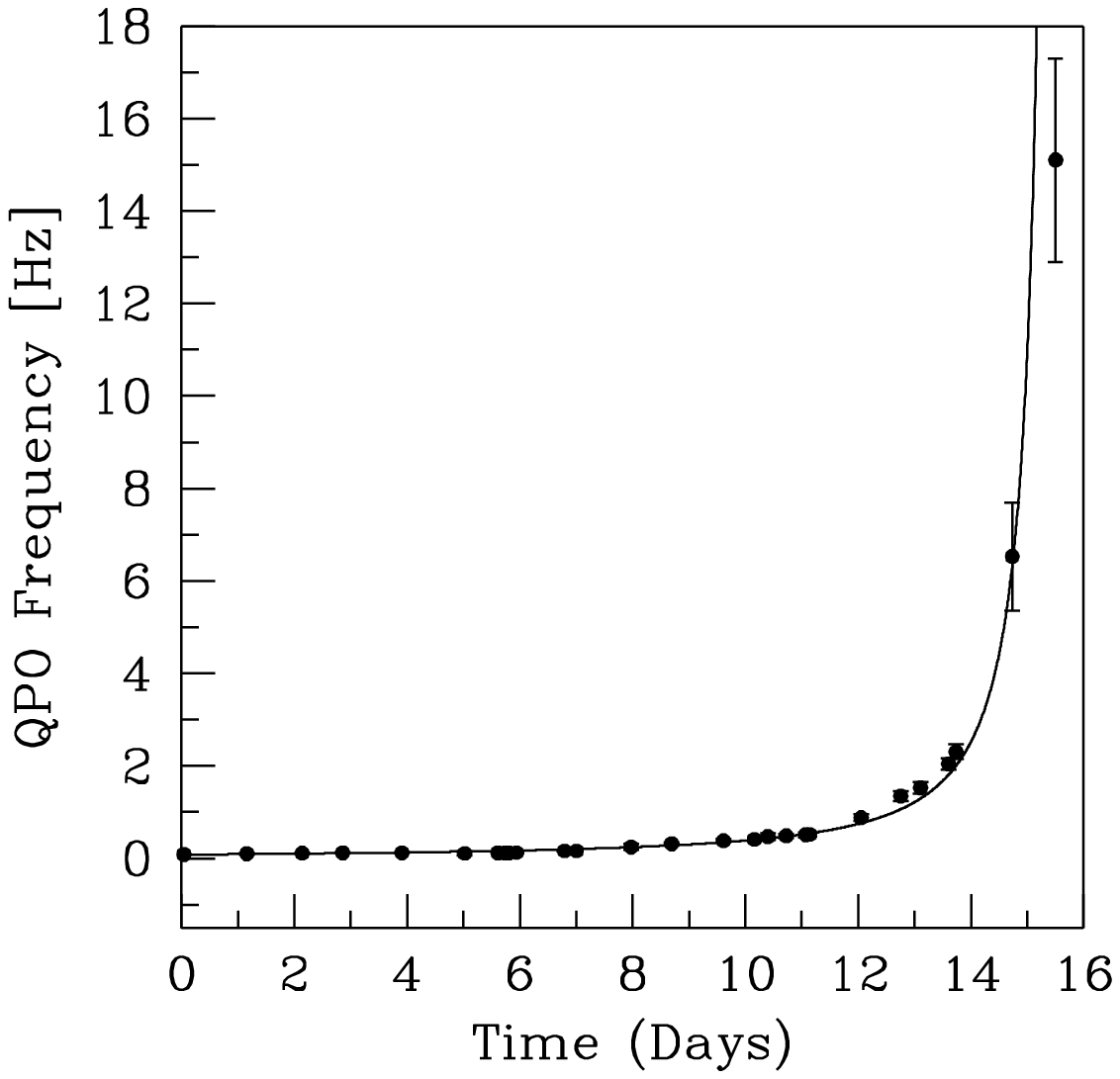,height=14truecm,width=14truecm}}}
\noindent {\small {\bf Fig. 4:}
Fig. 4: Observed centroid QPO frequencies (filled circles) of the first sixteen days
are compared with the model solution (solid curve) which assumes oscillatory shock propagating
towards the black hole with a  constant speed. Error-bars indicating
half line-widths obtained after fitting each power-density profile with
a power-law for the background and a Lorentzian profile for the QPO. The
final observation was made when the shock was at $97$ Schwarzschild radii from the
black hole.}

\end{figure}

\newpage

\begin{figure}
\vbox{
\vskip 0.0cm 
\centerline{ 
\psfig{figure=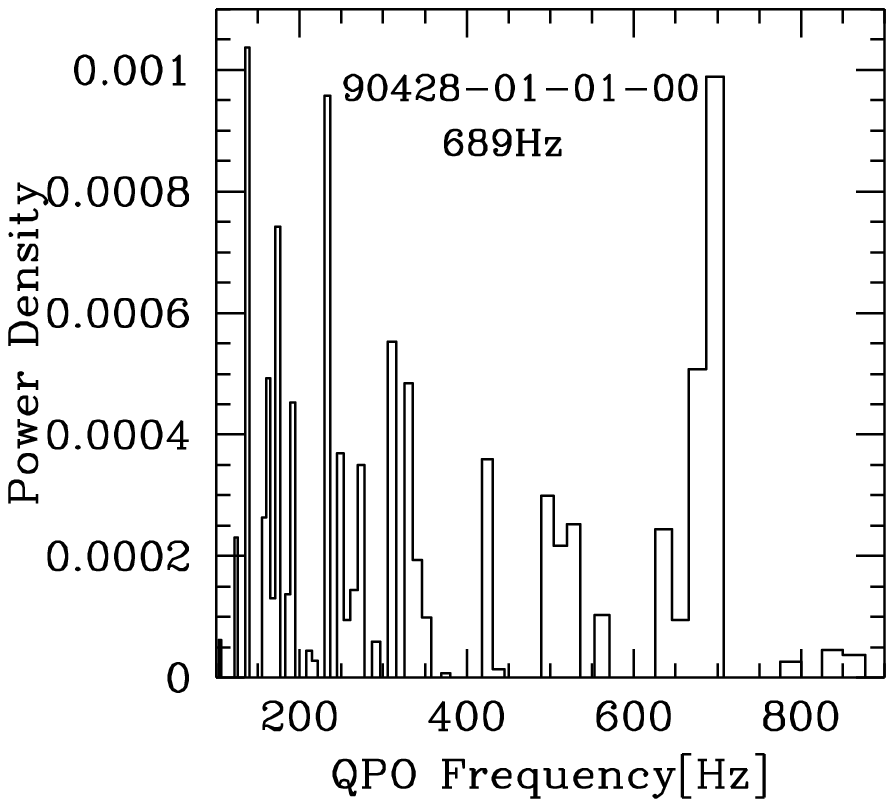,height=14truecm,width=14truecm}}}
\noindent {\small {\bf Fig. 5:}
Fig. 5: Possible signature of a high energy QPO in the current outbursting episode of GRO J1655-40.
In the power density spectrum presented, a QPO at $\nu_{QPO}\sim 689$Hz 
could be seen. This is the highest reported QPO for any black hole candidate in the literature. }

\end{figure}

\end{document}